\documentclass[aps,prl,reprint,groupedaddress,amsmath,showpacs]{revtex4-1}
\usepackage{scalefnt}
\usepackage{braket}
\usepackage{xfrac}
\usepackage{amssymb}
\usepackage{graphicx}
\usepackage{xcolor}
\usepackage{isotope}
\usepackage{pifont}
\usepackage{rotating} 
\usepackage{siunitx}
\usepackage{hyperref}
\usepackage{times,txfonts}

\newcommand{\symboldiamondsym}[1][black]{{\color{#1}\scalefont{0.8}\raisebox{-.2ex}{\begin{turn}{45}$\blacksquare$\end{turn}}}}
\newcommand{\symbolbox}[1][black]{{\color{#1}\scalefont{0.8}$\blacksquare$}}
\newcommand{\symbolcircle}[1][black]{{\color{#1}\scalefont{0.8}\ding{108}}}
\newcommand{\symbolcross}[1][black]{{\color{#1}\scalefont{0.8}{\ding{58}}}}

\definecolor{plot1}{rgb}{0.0,0.0,0.8}
\definecolor{plot2}{rgb}{0.8,0.0,0.0}
\definecolor{plot3}{rgb}{0.2, 0.7, 0.2}
\definecolor{plot4}{rgb}{0.61,0.32,0.61}

\begin{document}


\title{Importance-Truncated Large-Scale Shell Model}


\author{Christina Stumpf}
\email[]{christina.stumpf@physik.tu-darmstadt.de}
\author{Jonas Braun}
\author{Robert Roth}
\affiliation{Institut f\"ur Kernphysik, TU Darmstadt, Schlossgartenstr. 2, 64289 Darmstadt, Germany}


\date{\today}

\begin{abstract}

We propose an importance-truncation scheme for the large-scale nuclear shell model that extends its range of applicability to larger valence spaces and mid-shell nuclei. It is based on a perturbative measure for the importance of individual basis states that acts as an additional truncation for the many-body model space in which the eigenvalue problem of the Hamiltonian is solved numerically. Through \emph{a posteriori} extrapolations of all observables to vanishing importance threshold, the full shell-model results can be recovered. In addition to simple threshold extrapolations, we explore extrapolations based on the energy variance. We apply the importance-truncated shell model for the study of \isotope[56]{Ni} in the $pf$ valence space and of \isotope[60]{Zn} and \isotope[64]{Ge} in the $pfg_{\sfrac{9}{2}}$ space. We demonstrate the efficiency and accuracy of the approach, which pave the way for future shell-model calculations in larger valence spaces with valence-space interactions derived in \emph{ab initio} approaches.

\end{abstract}


\maketitle


\paragraph{Introduction.}

The nuclear valence-space shell model is one of the work horses in nuclear structure theory. It is very successful for the description of spectra and spectroscopic observables over a large range of nuclei and plays an important role in guiding and interpreting experiments from stable to  exotic nuclei \cite{Caurier2005, Horoi2006, Coraggio2009135, Otsuka2010, Lenzi:2010pe, Lenzi2010, Richter2011, Kaneko2011, Sieja2012, Poves2012, Holt:2010yb, Brown2013, Spinella:2013yza, Tsunoda2014, Coraggio2014, Langanke:2015opa, naidja2015recent}. Two critical aspects in the application of the shell model (SM) are the construction of the effective valence-space interaction as well as corresponding effective operators and the solution of the eigenvalue problem in the model space of the valence nucleons. 

Traditionally, effective valence-space interactions are constructed using renormalized nucleon-nucleon interactions combined with phenomenological fits of matrix elements to nuclei within the valence space \cite{Kuo1966, HjorthJensen1995125, Caurier2005, Coraggio2009135}. Though this phenomenological approach allows for a rather accurate description, it lacks a rigorous connection to the underlying nuclear interaction and does not provide a consistent framework for the treatment of observables other than the energy. Recently, a set of novel approaches to systematically derive valence-space interactions and operators have emerged \cite{Tsukiyama2012, Bogner2014, Jansen2014, Dikmen2015}. They offer new insights into valence-space interactions and can be linked to \emph{ab initio} calculations.  

Once the valence-space interaction is specified, the SM reduces to the solution of a large-scale matrix eigenvalue problem. Its dimension grows combinatorially with the number of valence orbitals and nucleons. Starting with valence spaces covering the $pf$-shell, the $m$-scheme model spaces reach dimensions beyond $10^{9}$ around mid-shell, which is approaching the limits of present computational approaches for sparse eigenvalue problems. When going to valence spaces covering more than one major shell, the model-space dimension poses a severe limitation to the applicability of the SM. In these cases additional truncations or more sophisticated methods like the Monte Carlo shell model (MCSM) \cite{Honma1995, Otsuka2001319} or a density matrix renormalization group treatment of the SM \cite{Legeza:2015fja} have to be employed. 

We propose the importance-truncated shell model (IT-SM) to overcome this limitation. It combines the SM with an importance-truncation scheme that is successfully applied in no-core configuration-interaction approaches for some time \cite{Roth2009}. In addition we use refined extrapolation schemes based on the energy variance to reduce the uncertainties of the IT-SM calculations. Together, importance truncation and extrapolation provide an accurate tool for systems and valence spaces beyond the reach of standard SM calculations.

\paragraph{Importance Truncation.}

The importance truncation is a physics-driven, adaptive truncation of the many-body model space based on a measure for the importance of individual basis states for the description of a specific set of eigenstates of a given Hamiltonian. The importance measure is defined through the amplitude of the individual basis states in the expansion of the eigenstates, obtained \emph{a priori} in lowest-order many-body perturbation theory. By imposing a threshold with respect to this importance measure we define an importance-truncated model space tailored specifically for the target eigenstates and Hamiltonian under consideration. Eventually, variations of the importance threshold and extrapolations to vanishing threshold can be used to extract observables in the limit of the full model space. This scheme is applied very successfully in the context of the no-core shell model (NCSM) \cite{Roth2007, Roth2009}.

The construction of the importance-truncated space is based on a set of reference states $\ket{\Psi_\text{ref}^{(m)}}$, which are obtained from a previous diagonalization in a small space, that represent the target eigenstates. The basis states that contribute to the reference states $\ket{\Psi_\text{ref}^{(m)}}$ span the reference space $\mathcal{M}_{\text{ref}}$. We estimate the importance of basis states $\ket{\Phi_\nu}$ outside $\mathcal{M}_{\text{ref}}$ by means of the amplitudes $\kappa_\nu^{(m)} = - \braket{\Phi_\nu|\,\mathbf{H}\,|\Psi_\text{ref}^{(m)}}/\Delta\epsilon_\nu$
of the first-order perturbative correction to $\ket{\Psi_\text{ref}}$, where the energy denominator $\Delta\epsilon_\nu$ corresponds to  the unperturbed single-particle excitation energy of the basis state $\ket{\Phi_\nu}$. Only basis states with importance measure $|\kappa_\nu^{(m)}|$ larger than a given importance threshold $\kappa_\text{min}$ for at least one reference state $\ket{\Psi_\text{ref}^{(m)}}$ are included in the importance-truncated model space.

In the case of a two-body Hamiltonian, the simple first-order importance measure cannot probe basis states that differ by more than a two-particle-two-hole (2p2h) excitation from any state in $\mathcal{M}_{\text{ref}}$. Therefore, we embed the construction of the importance-truncated space into an iterative scheme. For the valence-space SM, we use the number of valence particles above the orbits that are (partially) occupied in the lowest energy configurations to define a truncation parameter $T_{\max}$. For $T_{\max}=0$, all Slater determinants with valence nucleons distributed in the lowest accessible orbits are in the model space. For $T_{\max}=2$, up to two valence nucleons are promoted to higher-lying orbits---this model space can be generated through 1p1h and 2p2h excitations on top of the $T_{\max}=0$ space. Thus, we combine a sequential increase of the truncation parameter $T_{\max}$ with the importance-selection procedure. This sequential IT-SM scheme is analogous to the sequential IT-NCSM scheme discussed in Ref. \cite{Roth2009}.

The complete IT-SM calculation proceeds as follows: We start with a conventional SM calculation for small $T_{\max}$, e.g. $T_{\max}=0$, and select a set of target eigenstates. We define the reference states $\ket{\Psi_\text{ref}^{(m)}}$ by filtering the important components of these eigenstates through a so-called reference threshold $C_{\min}$ with respect to the amplitudes from the SM calculation. With these reference states we construct importance-truncated spaces with $T_{\max}=2$ for a sequence of importance thresholds $\kappa_{\min}$. In each space we solve the eigenvalue problem and compute the relevant observables. The eigenvectors for the largest importance-truncated space define the new reference states, again imposing a reference threshold $C_{\min}$, for constructing the importance-truncated spaces for $T_{\max}=4$. This procedure can be iterated until $T_{\max}$ reaches the number of valence particles and thus probes the full model space. In the limit $(\kappa_\text{min}, C_\text{min}) \rightarrow 0$, this algorithm is guaranteed to reproduce the results in the full model space at each $T_{\max}$.

\begin{figure}
\includegraphics[width=1\columnwidth]{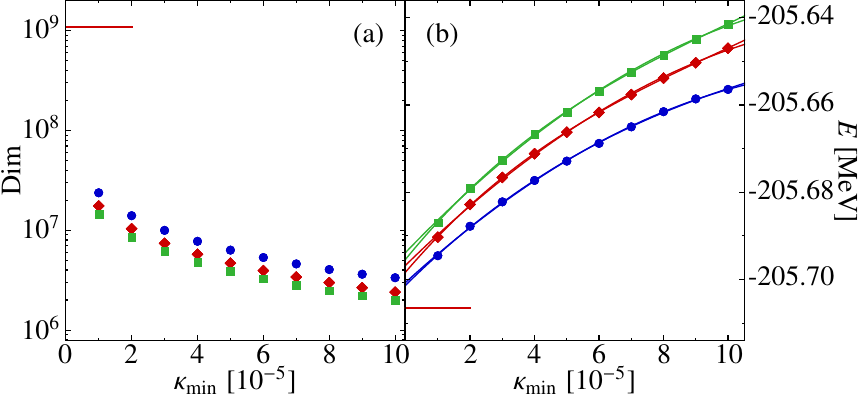}
\caption{\label{fig:fig1} (color online)
Dimension of the importance-truncated model space (a) and ground-state energy relative to the core (b) for \isotope[56]{Ni} in the $pf$ valence space as a function of the importance threshold for reference thresholds $C_{\text{min}}  = \{1\, $(\symbolcircle[plot1])$,\; 2\, $(\symboldiamondsym[plot2])$,\; 3\, $(\symbolbox[plot3])$\} \times 10^{-4}$ and $T_{\max}=16$ using the \textsc{gxpf1a} interaction.
The model space has been constructed for the simultaneous description of the six lowest eigenstates.
For the threshold extrapolation we use polynomials of order two and three.
The red lines denote the full $m$-scheme dimension and the ground-state energy of the full SM \cite{Horoi2006}.
}
\end{figure}

The results of IT-SM calculations for different thresholds $\kappa_{\min}$ and $C_{\min}$ are depicted in Fig.~\ref{fig:fig1}. As a test case, we consider \isotope[56]{Ni} in a $pf$ valence space using the \textsc{gxpf1a} interaction \cite{Honma2005}. The full $m$-scheme dimension of this model space is $1.09 \times 10^{9}$, which is at the limit of routine SM calculations. The results presented in Fig. \ref{fig:fig1} show the dimensions and the lowest energy eigenvalues as function of $\kappa_{\min}$. Note that the energy axis is extremely magnified and spans only 80 keV. The dimensions of the importance-truncated spaces are reduced drastically, by about two orders of magnitude as compared to the full SM space. At the same time, the absolute energies in the largest importance-truncated spaces, corresponding to the smallest $\kappa_{\min}$ and $C_{\min}$ thresholds, differ by only about 10 keV from the full SM. This demonstrates the efficiency of the importance truncation---it separates the $10^7$ basis states that determine the bulk of the energy from the $10^9$ basis states that are responsible for the residual 10 keV. 

\paragraph{Threshold Extrapolation.}

We can approximately account for the effects of basis configurations excluded from the importance-truncated spaces by an \emph{a posteriori} extrapolation of the observables. The simplest extrapolation addresses the importance threshold $\kappa_{\min}$. Since the energy eigenvalues depend smoothly on $\kappa_{\min}$ we can fit simple functions to the set of energies obtained for different $\kappa_{\min}$ values and extract the energies for $\kappa_\text{min} \rightarrow 0$. Since we do not have a theoretical model for the functional dependence on the importance thresholds, we use simple polynomials, typically of order two to four. Varying the order of the polynomials gives an estimate for the uncertainty of this threshold extrapolation. In Fig.~\ref{fig:fig1}(b) we have included examples for fits with second and third-order polynomials for the ground-state energies of \isotope[56]{Ni}. Note that the uncertainty of the $\kappa_{\min}$ extrapolation is small compared to the residual dependence on the reference threshold $C_{\min}$.  

This simple threshold extrapolation does not require additional computations and can be applied to all observables on equal footing (cf. Fig.~\ref{fig:fig5}). However, it exclusively addresses the importance threshold $\kappa_{\min}$ and uncertainties of the polynomial extrapolations can be sizeable. One can improve on this by including additional information on the excluded basis states, e.g., through a second-order perturbative estimate of their contribution to the energy, as done successfully in the IT-NCSM (see Ref. \cite{Roth2009} for details).

\paragraph{Variance Extrapolation.}

As an alternative to the simple threshold extrapolation, we consider a more elaborate extrapolation scheme based on the energy variance 
$\Delta E^2 = \braket{\Psi | \mathbf{H}^2 | \Psi} - \braket{\Psi | \mathbf{H} | \Psi}^2$, which was used in the SM context before \cite{Zhan2004, Shimizu2010, Shimizu2012}. By construction, the energy variance vanishes for the exact eigenstates and, thus, serves as a measure for the distance of an approximate state obtained in a truncated subspace from the energy eigenstate in the full space. As discussed in Ref.~\cite{Mizusaki2003}, the energy is expected to show a predominantly linear dependence on the energy variance, with sub-leading quadratic corrections. We thus have a simple model and a robust two- or three-parameter fit function at hand that provides accurate extrapolations.

The energy variance captures nontrivial information on the full model space through the expectation value $\braket{\Psi | \mathbf{H}^2 | \Psi}$. This is seen by inserting an identity operator represented in the full model space in between the product of the two Hamiltonians---the variance explicitly probes the coupling to states outside of the truncated subspace. In practical calculations we can choose the full space we wish to extrapolate to. In the context of the IT-SM it is desirable to extrapolate to the full SM space defined by the valence orbits without additional truncations. In this way the variance extrapolation remedies all truncations used in the IT-SM, i.e., the $\kappa_\text{min}, C_{\text{min}}$, and $T_{\text{max}}$ truncations. Therefore, the variance extrapolation is much more powerful than the simple threshold extrapolation.

\begin{figure}
\includegraphics[width=1\columnwidth]{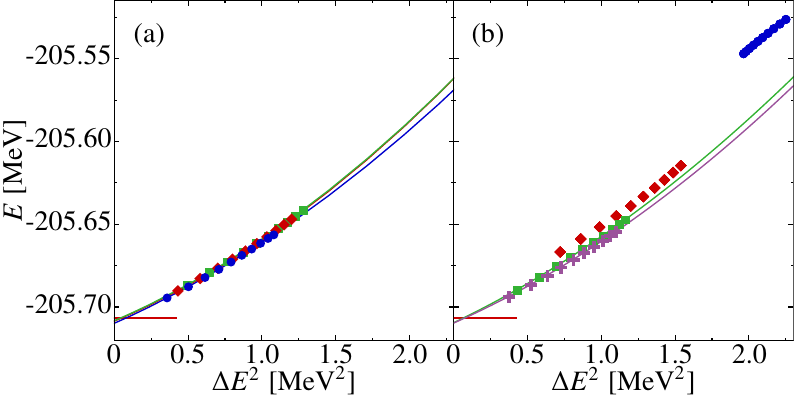}
\caption{\label{fig:fig2} (color online)
Energy-variance extrapolation of the ground-state energy relative to the core for \isotope[56]{Ni} obtained in IT-SM using the \textsc{gxpf1a} interaction.
In panel (a) results for different reference thresholds $C_{\text{min}}  = \{1\, $(\symbolcircle[plot1])$,\; 2\, $(\symboldiamondsym[plot2])$,\; 3\, $(\symbolbox[plot3])$\} \times 10^{-4}$ for $T_{\max}=16$ are shown. In panel (b) calculations for different truncations $T_{\max} = \{4 $(\symbolcircle[plot1])$,\; 6 $(\symboldiamondsym[plot2])$,\; 8 $(\symbolbox[plot3])$,\; 10 $(\symbolcross[plot4])$\}$ with $C_\text{min} = 1 \times 10^{-4}$ are depicted.
}
\end{figure}

In Fig.~\ref{fig:fig2} the variance extrapolation of the ground-state energy of \isotope[56]{Ni} is illustrated, where panel (a) shows $\kappa_{\min}$ sequences for different reference thresholds $C_{\min}$ and panel (b) shows $\kappa_{\min}$ sequences for different $T_{\max}$ truncations. The first remarkable observation is that the $\kappa_{\min}$ sequences for different $C_{\min}$ fall onto a straight line. Consequently the variance extrapolations for the different $C_{\min}$ give the same result. The variance-extrapolated energy is in excellent agreement with the result for the full space reported in Ref. \cite{Horoi2006}. Even with an additional $T_{\max}$ truncation, as shown in Fig.~\ref{fig:fig2}(b), the results beyond $T_{\max}=6$ fall onto the same line. For severe truncations, e.g. $T_{\max}=4$, we observe larger energy variances that cannot be extrapolated reliably. 

The advantages of the variance extrapolation are that a simple and robust fit model is available and that the extrapolation remedies all truncations inherent to an IT-SM calculation. The disadvantage is that substantial computational effort goes into the evaluation of the energy variance, typically the computation of the variance needs more computing time than the complete IT-SM calculation.

\paragraph{Applications.}

Using the IT-SM with threshold and variance extrapolation we now discuss the spectroscopy of \isotope[56]{Ni} in the $pf$ shell with the \textsc{gxpf1a} interaction \cite{Honma2005}. We demonstrate the robustness of the IT-SM by comparing energies and electromagnetic observables with full SM results obtained with the \textsc{Antoine} code \cite{Antoine,Caurier1999,Caurier2005} or extracted from Ref.~\cite{Horoi2006}.

\begin{figure}
\includegraphics[width=1\columnwidth]{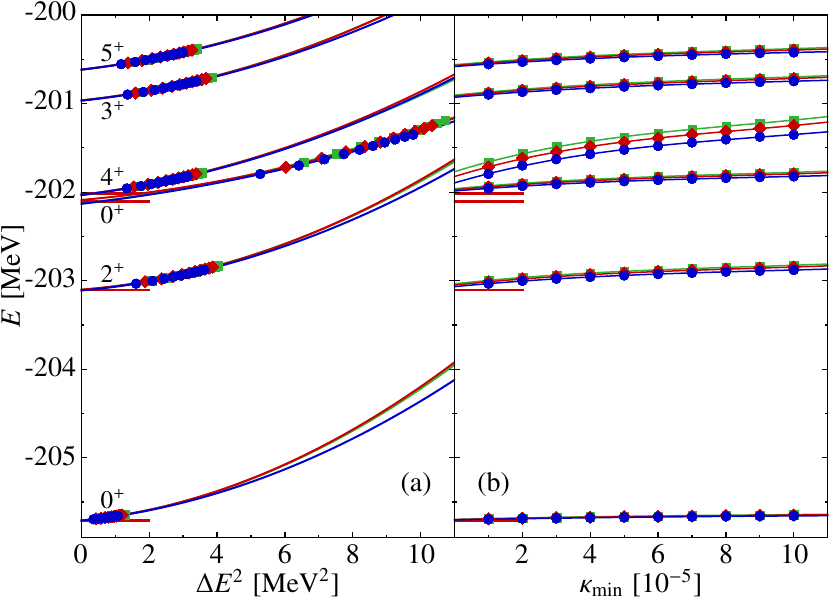}
\caption{\label{fig3}(color online)
Energy-variance (a) and threshold (b) extrapolation of the energies of the six lowest natural-parity states of \isotope[56]{Ni} using the \textsc{gxpf1a} interaction with $C_{\text{min}}  = \{1\, $(\symbolcircle[plot1])$,\, 2\, $(\symboldiamondsym[plot2])$,\, 3\, $(\symbolbox[plot3])$\} \times 10^{-4}$ and $T_{\max}=16$.
For the variance and threshold extrapolations, polynomials of order two and three have been employed, respectively.
The red lines show the full SM results extracted from \cite{Horoi2006}.
}
\end{figure}

Figure~\ref{fig3} shows the excitation spectrum of \isotope[56]{Ni} for three different reference thresholds as function of energy variance and importance threshold. On the scale of typical excitation energies the $\kappa_\text{min}$ and $C_{\text{min}}$ dependence is very weak. Both, the variance- and the threshold-extrapolated energies are in excellent agreement with the full SM results where available. The second $0^+$ state, however, shows a quite distinct behavior. Its $\kappa_{\min}$ and $C_{\min}$ dependences are stronger than for all other states and the energy variances are significantly larger. This indicates a particularly complicated structure, in this case due to deformation, resulting in many small components in the basis expansion of the eigenstate and, thus, a less accurate approximation in the importance-truncated space. The simple threshold extrapolation does not capture the contribution of all these small components and cannot correct for the sizeable $C_{\min}$ dependence. The variance extrapolation, however, provides a reliable extrapolation and even restores the correct level ordering in excellent agreement with the full SM. Particularly for these fragile states, the variance extrapolation offers significant advantages.

\begin{figure}
\includegraphics[width=0.8\columnwidth]{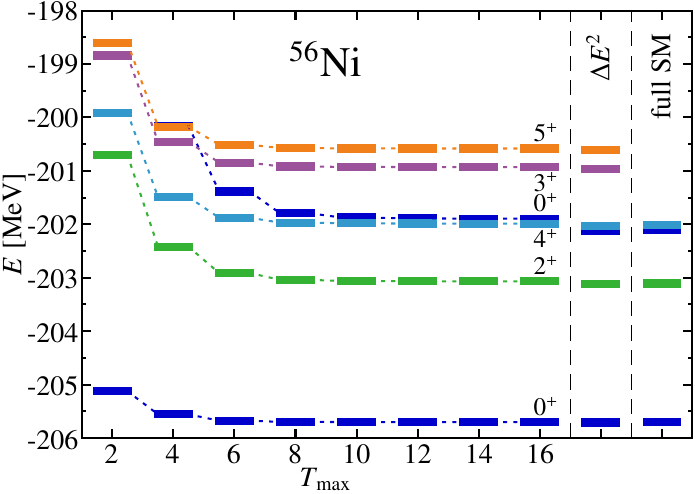}
\caption{\label{fig:fig4}(color online)
Natural-parity spectrum of \isotope[56]{Ni} as a function of $T_{\max}$ in terms of absolute energies relative to the core computed in the IT-SM with $C_\text{min} = 1 \times 10^{-4}$ using the \textsc{gxpf1a} interaction.
The right-hand columns show the results of an energy-variance extrapolation ($\Delta E^2$) and the full SM energies extracted from \cite{Horoi2006}.
}
\end{figure}

Figure \ref{fig:fig4} summarizes the extrapolated energies for the lowest six natural parity states of \isotope[56]{Ni}. The results of threshold extrapolations for a sequence of $T_{\max}$-truncated calculations are shown in the main part of the plot, followed by the spectrum obtained from the variance extrapolation with $T_{\max}=8$ and the full SM result \cite{Horoi2006}. Starting from $T_{\max}=8$ the spectrum is rather stable and in good agreement with the full SM results, except for the second $0^+$ state discussed above. The energy-variance extrapolation for $T_{\max}=8$ yields excellent agreement with the full SM for all states.

\begin{figure}
\includegraphics[width=1\columnwidth]{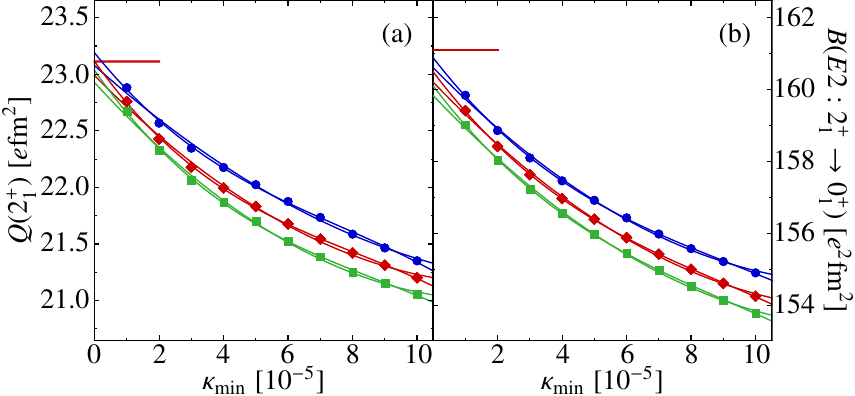}
\caption{\label{fig:fig5}(color online)
Threshold dependence and extrapolation for the quadrupole moment of the $2^+_1$ state (a) and the $B(\text{E2}: 2^+_1 \rightarrow 0^+_1)$ transition strength (b) for \isotope[56]{Ni}. The wave functions have been obtained in an IT-SM calculation using the \textsc{gxpf1a} interaction for $T_{\max}=8$ and the reference thresholds $C_{\text{min}}  = \{1\, $(\symbolcircle[plot1])$,\, 2\, $(\symboldiamondsym[plot2])$,\, 3\, $(\symbolbox[plot3])$\} \times 10^{-4}$. The red lines represent the full SM results obtained with the \textsc{Antoine} code \cite{Antoine, Caurier1999, Caurier2005}.
}
\end{figure}

Since the IT-SM also provides the eigenstates in the importance-truncated space, we have access to all other observables, particularly to electromagnetic moments and transitions relevant for spectroscopy. For each $\kappa_{\min}$ we compute the observable of interest using the respective eigenvector. Figure \ref{fig:fig5} illustrates the dependence of the quadrupole moment and the $B(E2)$ transition strength from the first $2^+$ state to the ground state in \isotope[56]{Ni} on the importance threshold. Also these observables show a smooth dependence on $\kappa_{\min}$ and allow for simple polynomial extrapolations to vanishing importance threshold. There is a mild dependence of the $\kappa_{\min}$-extrapolated results on $C_{\min}$ which is of the same magnitude as the uncertainty of the $\kappa_{\min}$ extrapolation. Within these small uncertainties, the extrapolated quadrupole moment and $B(E2)$ transition strength are in excellent agreement with full SM calculations proving that spectroscopic observables are also directly accessible in the IT-SM. We note that extrapolations using the energy variance do not improve these results.

We conclude this discussion with a first application of the IT-SM in a valence space covering more than one major shell. This will be an important future field of application of the IT-SM in conjunction with the new valence-space interactions derived in \emph{ab initio} approaches. We consider a $pfg_{\sfrac{9}{2}}$ valence space using the \textsc{pfg9b3} interaction \cite{PFG9B3ref1, PFG9B3ref2} and study \isotope[60]{Zn} and \isotope[64]{Ge} with full model-space dimensions of $2.2 \times 10^{13}$ and $1.7 \times 10^{14}$, respectively. Since these extended model spaces are susceptible to center-of-mass spuriosities, we use a Lawson prescription to diagnose center-of-mass contaminations \cite{Gloeckner1974313}. 
Particularly \isotope[64]{Ge} has been studied before in the MCSM \cite{Shimizu2010, Shimizu2012} using the same interaction. Very recently, a study using the density-matrix renormalization group with the SM targeted the same nucleus and valence space \cite{Legeza:2015fja}. These competing approaches highlight the difficulty of these calculations.

\begin{figure}
\includegraphics[width=1\columnwidth]{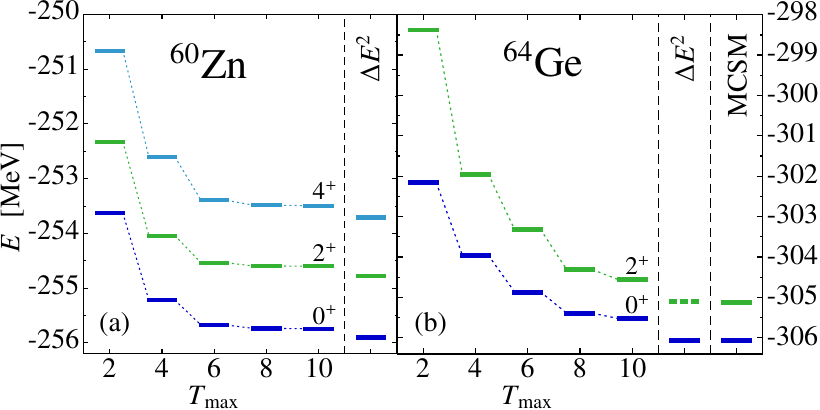}
\caption{\label{fig:fig6}(color online)
Lowest natural-parity states of \isotope[60]{Zn} (a) and \isotope[64]{Ge} (b) computed in the IT-SM for $C_{\text{min}} = 2 \times 10^{-4}$ using the \textsc{pfg9b3} interaction with subsequent threshold extrapolation for different values of $T_{\max}$.
The right-hand columns show the results obtained from the energy-variance extrapolation ($\Delta E^2$) of the $T_{\max}=6$ results.
The dashed line shows an approximation for the energy of the $2^+$ state calculated from the excitation energy obtained in the IT-SM for $T_{\max}=10$ and the $\Delta E^2$-extrapolated ground-state energy.
For \isotope[64]{Ge}, the MCSM results \cite{Shimizu2010,Shimizu2012} are shown for comparison.
}
\end{figure}

Figure \ref{fig:fig6} shows the absolute energies of the lowest states in \isotope[60]{Zn} and \isotope[64]{Ge} extracted from a simple threshold extrapolation for a sequence of $T_{\max}$ truncated spaces and from an energy-variance extrapolation. Whereas the spectra seem converged at $T_{\max}=8$ for \isotope[60]{Zn} there is still some dependence on  $T_{\max}$ for \isotope[64]{Ge}. Moreover, for \isotope[64]{Ge} the variance extrapolation gives the ground-state energy about 0.5 MeV lower than the threshold-extrapolated energy at $T_{\max}=10$, due to effects of the $C_{\max}$ and $T_{\max}$ truncations ignored in the threshold extrapolations. The sensitivity to these truncations results from the strong deformation of \isotope[64]{Ge}, which requires many small components in the SM basis expansion to describe. The variance extrapolation captures these subtle effects and yields excellent agreement with the MCSM results \cite{Shimizu2010,Shimizu2012}.

\paragraph{Conclusions.}

We have introduced the IT-SM approach and demonstrated its ability to extend the reach of valence-space SM calculations into the domain of large valence spaces and mid-shell nuclei. In addition to the threshold extrapolation, we adopted an extrapolation in terms of the energy variance for the first time in the IT context. Generally, the threshold extrapolation provides sufficiently accurate energies and electromagnetic observables at no extra computational cost. In specific cases, e.g. for states governed by deformation, the energy-variance extrapolation provides better accuracy for energies at significant extra cost.
 
The IT-SM framework is ideally suited to study valence spaces spanning two or more major shells with effective interactions derived in an \emph{ab initio} framework, such as the in-medium similarity renomalization group \cite{Tsukiyama2012, Bogner2014} or the Lee-Suzuki approach \cite{Jansen2014, Dikmen2015}. Together, these new developments offer unique perspectives for detailed nuclear structure investigations beyond the reach of the conventional SM.

\begin{acknowledgments}
We thank G. Mart\'inez-Pinedo for useful discussions and T. Otsuka for providing us with the \textsc{pfg9b3} interaction.
This work is supported by the DFG through contract SFB 634, the Helmholtz International Center for FAIR (HIC for FAIR), and the BMBF through contract 06DA7047I.
The authors gratefully acknowledge computing time granted by the CSC Frankfurt (LOEWE-CSC) and the computing center of the TU Darmstadt (LICHTENBERG).
\end{acknowledgments}

%

\end{document}